# Nodal cooperation equilibrium analysis in multi-hop wireless ad hoc networks with a reputation system

Jerzy Konorski    Karol Rydzewski

jekon@eti.pg.gda.pl    k.rydzewski@o2.pl

Faculty of Electronics, Telecommunications and Informatics
Gdansk University of Technology
Gdansk, Poland

**Abstract.** Motivated by the concerns of cooperation security, this work examines selected principles of state-of-the-art reputation systems for multi-hop ad hoc networks and their impact upon optimal strategies for rational nodes. An analytic framework is proposed and used for identification of effective cooperation-enforcement schemes. It is pointed out that optimum rather than high reputation can be expected to be sought by rational nodes.

**Keywords:** ad hoc network, cooperative security, utility, cost, reputation system, cooperation, strategy

## 1    Introduction

Wireless ad hoc networks are gaining popularity as a growing number of areas of human activity require omnipresent and self-configuring connectivity. One of the greatest advantages of this network concept – the lack of central infrastructure and governing authority – is its greatest challenge as well. Until its inception, standards ruling networks' operation were obeyed ensuring their optimal performance. In the ad hoc paradigm, this fundamental law is disputed, as the autonomy of network nodes enables them to neglect collective welfare and selfishly optimize their individual utility [1]. Motivations behind these decisions may be of different origin, to boost node's performance, lengthen operational life, etc. Some of these effects may be introduced intentionally by node's software programmers, others may emerge out of non-hostile optimization techniques employed. The scarcity of resources, namely of battery power and bandwidth, aggravate the problem.

Selfishness ensures a better performance of a node compared to nodes following the primarily altruistic network standards. The impact on the network is minimal until there are few selfish nodes. However, other nodes in the network may quickly learn to acquire the same strategy. As the number of selfish nodes grows, the network performance drastically deteriorates and eventually the network disintegrates, calling into question its very mission. This is clearly a security (more precisely, cooperation security) issue, since it directly impacts the inter-nodal communication capability.

Such considerations are not unusual in the realm of game theory [2]. They are used to explain a well-known phenomenon in multi-hop networks, where players following their best interest choose strategies resulting in non-optimal solutions for everyone, including themselves. In such games, the only solution improving the game results seems to consist in shaping the game in such a way that the non-cooperative (selfish) strategies become non-optimal for the players, and in promoting cooperative ones.

The idea of discouraging network nodes from becoming selfish underlies the idea of *reputation systems*. Such a system, fed with experience reports from nodes that have had interactions with other nodes, disseminates information that helps deciding whether a certain node can be trusted to deliver a service on a certain service level. Other nodes, having this information can make decisions whether or not to cooperate with such a node, without any prior experience with it. This leads to the economic concept of *indirect reciprocity* [3], where nodes are rewarded or punished for their prior cooperative or noncooperative behavior towards nodes different from the ones they are currently interacting with. Such a system, to be useful and to actually encourage nodes to cooperate, has to be carefully designed to reshape the payoff matrix of the arising game so that globally non-optimal behavior becomes non-optimal to individual nodes as well.

In this paper we examine some state-of-the art principles of reputation systems with a special interest on how they influence individual nodal strategies and how these strategies influence network operations, in particular if they actually discourage undesirable behaviors. This paper is structured as follows: in Section 2 we briefly present the state of the art in reputation systems and previous work; in Section 3 used metrics are discussed; Section 4 discusses reputation systems' typical design assumptions and their impact on nodal strategies; Section 5 examines possible nodal responses to modifications of utility functions; finally, Section 6 summarizes our findings and concludes the paper.

## 2    Related work

In recent years, reputation systems have become a popular research avenue addressing the problem of selfish nodes and cooperation enforcement. The basic concept entails a method of monitoring nodes' behavior, a behavior-rating algorithm, reputation calculation and an algorithm of leveraging the calculated reputation in network operations. The key part of a reputation system is the cooperation detection and evaluation. The *watchdog* mechanism [5-10] is widely used in numerous wireless network environments. Its principle of work is based on omnidirectional characteristics of antennas typically employed by wireless network nodes. Assuming this, a node's neighbors that are situated within the radio range can overhear all its communication. This mechanism is, however, innately unable to address non-uniform radio range, transmission impairments, and unpredictable collisions. Other solutions focusing on identifying a single misbehaving network node are based on a subnetwork of cooperating nodes observing the environment in their proximity and sending reports to other nodes. Such approach is exemplified by the Two-ACK scheme [5], based on addi-tional short range ACKs sent by the intermediate nodes on a given path, and on the flow conservation presumption [6]. According to it, cooperative intermediate nodes keep a count of transit traffic and share it with other nodes; thus they are able to identify nodes responsible for "leaking" packets. A different concept, deriving agent reputation from composed service of multiple agents [7], is used in [8] for detection of selfish nodes in military wireless sensor networks of a hierarchical directed tree topology. The detection and avoidance of malicious nodes is orchestrated by a sink node. The operation of this system is divided into rounds, in each of which the sink node changes the network topology. The sink node gathers statistics on network behavior. Based on the delivery ratio for packets sent by a given source node, as well as the network topology and its changes, the sink is able to deduce malicious and suspicious nodes in the network. A similar concept [9], based on end-to-end ACKs is used to deduce behavior of intermediate nodes from multiple reports on different paths in a MANET network; certainty levels of the results can be computed.

Based on the results obtained from various detection mechanisms, usually in the form of delivery or forwarding ratios, nodes' reputation is calculated. Many reputation systems serve just to discern cooperative or misbehaving nodes measuring their behavior against a predefined desirable pattern [4] and returning a binary value describing a node's positive or negative rating. Sometimes this binary metric is extended to define intermediate states or to indicate nodes of uncertain reputation. Alternatively, a more fine-grained view of a node is created [9], typically with real-valued reputation levels between 0 and 1, where 0 represents a completely uncooperative node and 1 a fully cooperative one. Usually reputation reflects nodes' behavior in a certain period of time and may be constantly recalculated. In some systems, when a node is labeled as uncooperative, it is regarded as such forever, whereas other systems offer possibilities to regain positive, cooperative rating, either by resetting it after a predefined period of time or offering a "redemption" opportunity.

The use of the calculated reputation can be twofold. On the service requesting side it is often used as an extension to routing algorithms assisting in path selection. A well-known solution, called *pathrater* [4], attempts to select a path without misbehaving nodes. If such a path is unknown to exist at the moment, a path rediscovery phase is triggered. However, some authors, e.g., [10], stress that routing around misbehaving nodes in fact rewards them, as they incur lower costs of participating in the network; accordingly, they propose to route part of the traffic via lower-reputation nodes. This gives a node a possibility to regain ("redeem") high reputation and diversifies traffic among multiple nodes, while discouraging whitewashing (i.e., changing identity to restore unblemished reputation). On the service provision side, specifically when the service consists in forwarding transit packets, punishment may be administered by only forwarding packets originated at a cooperative source node, or doing so with a probability depending on the source node's reputation level [10]. Some systems use only the service requesting side, which may actually promote misbehaving nodes, giving a selfish strategy so called *evolutionary stability*. Recently, this concept, well-known in game theory and in biology [3], has been linked to reputation systems to provide a more in-depth explanation of indirect reciprocity. In [3], a vast number of possible behaviors (or strategies) coupled with reputation systems are critically exam-ined in order to identify evolutionarily stable (strategy, reputation system) pairs that perform well against nonstandard alien behavior. Best performing pairs are found to adhere to similar principles: by giving help to a good agent, the donor also earns a good reputation, whereas refusal of help to a good agent brings a bad reputation; refusing to help ill-reputed agents does not undermine a good reputation. The authors note that ensuring evolutionary stability against a group of alien agents entails regarding helping a good agent as good, and helping a bad agent as bad. Thus indirect reciprocity, i.e., cooperative behavior toward third parties, is promoted.

In [11], a game-theoretic study of nodes' behavior dynamics is presented and an attempt is made to search an equilibrium. The authors propose a reputation system where nodes locally observe the behavior of paths they are using based on their delivery and deduce therefrom the behavior of every node. The authors associate the gain of a node with sends a successfully delivered source packet or receives a destination packet; the loss is associated with forwarding a transit packet on behalf of some other source node. The parameters under nodes' control are: the forwarding ratio of transit traffic on a given incoming link, the amount of outgoing traffic and a threshold value of packets dropped by other nodes on a given outgoing link. Based on those arbitrarily set values and a utility function, a node is able to adjust its behavior towards other nodes, which in extreme cases may even result in completely shutting a given network link. The influence of the network size and traffic volume is demonstrated – the more source traffic a node has, the better forwarding service it provides.

## 3   Network resources and utility functions

We restrict the focus of a reputation system to packet forwarding along paths set up as sequences of nodes to traverse. However, our reasoning can be carried over to other types of network service. The service can be measured and quantified into a set of key performance indicators (KPIs), enabling evaluation and comparison of the implemented solutions. In the history of computer networks, a significant number of KPIs have been developed. A network node requests a specific amount of service to be provided by the network that it may regard as an abstract external entity. Due to different network phenomena, limited communication capacity, protocol constraints or node's policy, the node's requests can be serviced either in full or to some extent, or completely rejected. Each unit of service costs the requesting node and the servicing entity some defined cost, which we model as constant across all the nodes. Cost-based analysis of nodes' and network's operations is a well-established direction of research in computer networks. To discover and evaluate nodal motivations behind selfishness one needs to know what drives their behavior and how their costs (also referred to as *utilities*) are created. This knowledge allows one to build effective network solutions shaping nodal utility in a way that is desired from the network perspective.

To examine nodes' strategies in a simplified model we propose a concept of a game between a node *X* and the network *N*, pictured in Fig. 1. The two parties in this game are issuing towards each other service requests and are interested in achieving optimal performance. Hence, both employ techniques enabling them to evaluate an-other party behavior and to respond to it adequately. The network $N$ employs a reputation system and utilizes the concept of indirect reciprocity, and is modeled by node $X$ as a single, albeit complex entity. On the technical level, indirect reciprocity is ensured by an agreed upon (centralized or distributed) algorithm of reputation calculation, dissemination the reputation values among network nodes, and a specific algorithm of nodal response to the assigned reputation. There are many methods employed to resolve these issues, some of them were summarized in the related work section. Node $X$ employs the concept of rationality and ability to define its own cost-effective strategies. In the presented game model, following variables and parameters are taken into consideration:

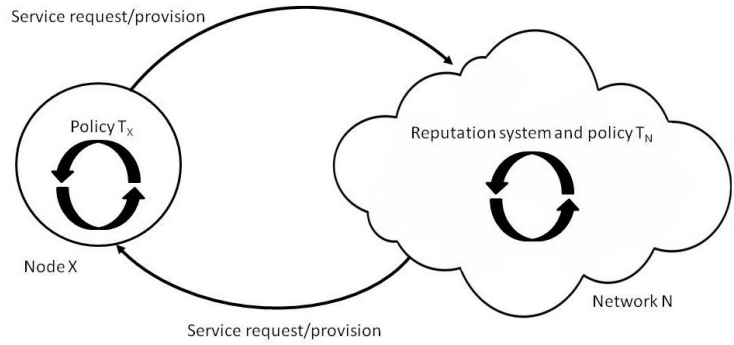

**Fig. 1.** Model of a game between node $X$ and the network $N$.

- $S_{XN}$ – amount of service (in service units) requested by node $X$ from the network $N$.
- $S_{NX}$ – amount of service requested by the network $N$ from node $X$; in our model, for the sake of meaningful evaluation of node $X$'s behavior, $S_{NX} > 0$ is assumed.
- $T_X \in [0, 1]$ – policy of node $X$ stating what proportion of $S_{NX}$ is to be provided,
- $U_X$ – utility node $X$ draws out of being connected to the network,
- $b$ – available bandwidth (in service units),
- $e$ – observation error reflecting the inability of the network $N$ to correctly assess the behavior of node $X$ because of the flaws in the network operation and/or radio environment,
- $R_X \in [0, 1]$ – reputation of node $X$ related to the value of $T_X$ evaluated by $N$ in a way defined by chosen reputation system, i.e., $R_X$ is a function of $T_X$, $S_{NX}$ and $e$,
- $T_N \in [0, 1]$ – policy of the network $N$ towards node $X$, i.e., the proportion of $X$'s requested service $S_{XN}$ that is provided; it is a function of $R_X$, and
- $G$ – service value; a constant representing how valuable for node $X$ is the service it requested from the network $N$, i.e., how much node $X$ is determined to pay for $S_{XN}$ if it is serviced by the network $N$.

The above variables and parameters result from the network design decisions, (in particular concerning the reputation system the indirect reciprocity mechanism), the radio environment, and the network's and node $X$'s requirements for service. The policy $T_X$ is the only variable that can be set according to node $X$'s strategy. Because

$T_X$ is directly unobservable, the network can only rely on $R_X$ calculated by the reputation system based on the past observations of node $X$'s behavior. Ideally, if the system uses a fine-grained reputation metric. If $e = 0$ and the observations sample is large enough, $R_X$ should reflect $T_X$ precisely for the network $N$. We will use the assumption $R_X = T_X$ henceforth.

To evaluate node $X$'s utility we propose a utility function (1), representing node $X$'s gain from being connected to the network. The goal of node $X$ is to select a strategy that maximizes its utility function with respect to the chosen service, i.e., forwarding source traffic, while taking some environmental limitations into account (2). Hence, the optimization problem for node $X$ can be stated as follows:

$$U_X = f(G, S_{XN}, T_N) \qquad (1)$$

$$T_X^* = \arg\max_{T_X \in [0,1]} U_X \left| S_{NX}(1+T_X) + \frac{S_{XN}}{T_N} \leq b \right. \qquad (2)$$

$U_X$, is a function of node $X$'s behavior, with respect to the amount of service it provides to the network $N$, as assessed by the network's reputation mechanism producing the value $R_X$, and the amount of service the node $X$ receives from the network $N$. Node $X$, being in control only of its service policy, in response to the obtained $U_X$, assesses its optimal $T_X$ to maximize its utility, taking into account the network's limited bandwidth for service provision (2). The bandwidth is consumed by issuing own requests taking into account requests that need to be reissued (i.e., $S_{XN}/T_N$), and processing network's requests (i.e., $S_{NX}(1+T_X)$, where the $(1+T_X)$ term comes from the nature of wireless network operation: $S_{NX}$, even if not processed by node $X$, consumes its bandwidth, since received requests occupy the radio medium and nothing can be broadcast at that time). Note that this model assumes that the rejected service requests should be reissued by node $X$ until they are successfully processed by the network. In reality some threshold on number of reissues should be implemented to avoid a negative $U_X$. However, the aim of this work is to show the effects of node $X$'s rational behavior, and introducing such thresholds could blur the overall picture.

The above game in which node $X$ optimizes its utility against the employed reputation system, and the network attempts to accomplish its mission of providing network-wide connectivity and well-being continues in search of an equilibrium. The general problem (1) takes on various forms depending on the specific network services and operations, in particular the choice of the workings of the reputation system and the algorithm setting $T_N$ in response to $R_X$. Examples of these specific formulations will be examined in Section 4.

## 4 Reputation systems' design impact on utility function

There exist various methods of monitoring nodes' behavior as detailed in Section 2; most of them are not free from flaws that prevent them from determining nodes' behavior precisely. The complexity of wireless network operation as well as the diversi-ty of possible factors influencing it make it impossible to eliminate those flaws completely. One way of dealing with this is to accept the measurements' imperfections and account for them when modeling the network's and reputation system's operation. This is the approach we follow here by including the *e* variable reflecting these flaws. Reputation, as an algorithm transforming observed node's behavior into numerical values of reputation, also has an impact upon the network's ability to distinguish and address various nodal behaviors.

The most important part of the reputation system, from the operational cost viewpoint, is the indirect reciprocity mechanism mentioned earlier, i.e., a set of algorithms constituting part of the network operation that reflect node *X's* reputation. Indirect reciprocity may transform a reputation system from a pure signaling tool into an enforcement mechanism fostering nodes' cooperative behavior. In what follows we examine common design concepts and their influence on node *X*'s utility function.

### 4.1 Plain multi-hop ad hoc network

A plain multi-hop ad hoc network does not account for nodes' rationality and assumes that all the nodes follow some predefined standard behavior. In such a network, there is no reputation system and the network provides service in response to node *X*'s service requests under a best effort policy regardless of its (not even monitored) behavior. This is equivalent of $R_X \equiv 1$, and the utility function (1) and node *X*'s policy (2) are expressed as in (3) and (4), respectively:

$$U_X = (G-1) \cdot S_{XN} - T_X \cdot S_{NX} \tag{3}$$

$$T_X^* = \arg\max_{T_X \in [0,1]} U_X \bigg| S_{NX}(1+T_X) + S_{XN} \leq b \tag{4}$$

*X*'s utility (3) is the difference between the gain from having own service requests successfully serviced (($G-1) \cdot S_X$) and the cost associated with servicing the network's requests ($T_X \cdot S_{NX}$). We subtract 1 from *G* to reflect the cost associated with issuing the amount $S_{XN}$ of requests, which lowers node *X*'s gain.

### 4.2 Tit-for-tat reciprocity mechanism

A tit-for-tat-type reciprocity mechanism enables the network *N* to respond in kind to node *X*'s behavior, i.e., *N* provides requested service to *X* in the same proportion that *X* provides requested service to *N*: $T_N = T_X$. Formulas (1) and (2) transform into (5) and (6), respectively:

$$U_X = G \cdot S_{XN} - S_{XN}/T_N - T_X \cdot S_{NX} \tag{5}$$

$$T_X^* = \arg\max_{T_X \in [0,1]} U_X \bigg| S_{NX}(1+T_X) + S_{XN} + \frac{S_{XN}}{T_N} \leq b \tag{6}$$

Formula (5) contains a term representing node $X$'s service requests rejected by the network that need to be reissued ($S_{XN}/T_N$), which was already discussed in (2). This replaces subtracting 1 from $G$ in (3).

### 4.3 Reputation metric

Reputation metrics in general need not impact the mechanisms of monitoring node $X$'s behavior. As stated earlier, some common reputation metric types are fine-grained and binary. The former takes node $X$'s observed behavior to be numerically equal to its reputation, optionally with some additional scaling to reflect all possible behaviors. The latter incorporates an algorithm of transforming all possible behaviors into a discrete, two-valued metric. Typically, this algorithm imposes a threshold $t_S$ upon the ratio of $S_{XN}$ and $S_{NX}$, the crossing of which changes $R_X$. The type of reputation metric does not directly influence formulas (1) and (2), as it only influences the shape of the function $R_X(T_X, S_{NX}, e)$.

### 4.4 $S_{NX}$ reflecting node $X$'s reputation

Besides enabling tit-for-tat service provision, reputation is often meant to assist network nodes in selecting appropriate (trustworthy enough) nodes to interact with and request service from. The basic approach dictates that only highest-reputed nodes be considered. Other approaches are more refined and make use of lower-reputed nodes as well, allowing non-optimal performance.

In the latter case, irrespective of the actual policy $T_N$, it is often assumed that the network $N$ is able to split its service requests to be alternatively serviced by nodes other than node $X$, hence node $X$ gets only a part of the original amount of requested service $S_{NX}$. To achieve this, the network $N$ requires an algorithm to transform $R_X$ into a parameter determining the part of the original $S_{NX}$ directed to the node $X$. For simplicity, we will assume this parameter is equal to $R_X$ (a general derivation of an optimal transformation of $R_X$ into this parameter is beyond the scope of this paper). Formulas (1) and (2) change into (7) and (8), respectively:

$$U_X = (G-1) \cdot S_{XN} - R_X T_X S_{NX} \qquad (7)$$

$$T_X^* = \arg\max_{T_X \in [0,1]} U_X \bigg|_{S_{NX}(1+T_X)R_X + S_{XN} \leq b} \qquad (8)$$

Equations (7) and (8) introduce $R_X$ as a parameter influencing the amount of service requested by the network $N$ from node $X$. The other terms are as in (3) and (4).

## 5 Rational nodes' strategies and their effects

In this section we examine the concepts presented in Section 4, either by themselves or in selected combinations employed in the state-of-the-art reputation systems. We will focus on the impact of a given solution on node $X$'s utility and available strategies. We will anticipate how the behavior of node $X$ impacts the network operation and check if a given solution encourages nodes to become cooperative. One of our goals in this analysis will be how the utility $U_X$ and optimal $T_X$ for node $X$ change depending on the ratio $M$ of the requested amounts of service:

$$M = \frac{S_{NX}}{S_{XN}} \qquad (9)$$

### 5.1 Plain multi-hop ad hoc network

In a network without a reputation system, the utility function $U_X$ given by (1) has two components – one representing service requests originating from node $X$, and another one representing the cost of providing service in response to the network $N$'s requests. The policy $T_X$, the only part of the model controllable by node $X$ and acting upon $S_{NX}$, is bounded only by the above cost and has no bearing upon $S_{XN}$. The dominant strategy of the node $X$ in this situation is trivially $T_X = 0$, irrespective of any parameters defining the game, meaning a completely uncooperative node. Therefore, if every network node were to be rational in terms of costs and gains, and follow this strategy, the network would disintegrate.

### 5.2 $S_{NX}$ reflecting $X$'s reputation

Solutions using this concept have only one possibility of influencing forwarding decisions, which is by shaping the amount of service $S_{NX}$ the network $N$ requests from node $X$ in step with $R_X$, and in this way shaping node $X$'s utility. The incentives indicated in Section 5.1 become even stronger in equation (7), since $S_{NX}$ typically increases as $R_X$ increases. However, $R_X$ is mainly influenced by $T_X$, therefore by staying uncooperative and so keeping a low reputation, node $X$ can lessen the amount $S_{NX}$. Thus its optimal policy is $T_X = 0$. Apart from a high utility, the optimal strategy also ensures a greater amount of available bandwidth, as transit traffic avoids the node $X$. A frequently used variant of this solution introduces a threshold $t_P$ on $R_X$ below which no service is requested from node $X$; at the same time, no reciprocity mechanism is used. The optimal $T_X$ changes and is now anywhere in $[0,t_P)$. However, from the network $N$'s perspective, it produces the same effect as in the variant without the threshold implemented, since no service is provided by node $X$.

### 5.3 Tit-for-tat reciprocity mechanism with fine-grained reputation metric

In this subsection we analyze equations (5) and (6) extended with $R_X$ taken from (7) and (8) along with fine-grained reputation metric detailed in Sec. 4.3. The node $X$'s optimal policy becomes more complex when playing against a network using a tit-for-tat fine-grained strategy, and becomes then more dependent on $M$. If $M < 1$, the optimal policy for $X$ is $T_X = 1$. However, as $M$ grows, the optimal $T_X$ decreases though at much slower rate (Fig. 2.), i.e., is roughly proportional to the logarithm of *M*. The utility at $T_X = 0$ is always infinite negative, i.e., no service is provided to *X* by the network *N*. In all cases charted on Fig. 2 constant *G* equals 10. However, as it can be observed on the dotted line chart, representing the case were *M*=50, value *G* is insufficient to elevate $U_X$ above *0*. A rational node, in this case, should restrain from providing any service to the network, i.e. $T_X=0$ as well as it should stop issuing its own request to the network.

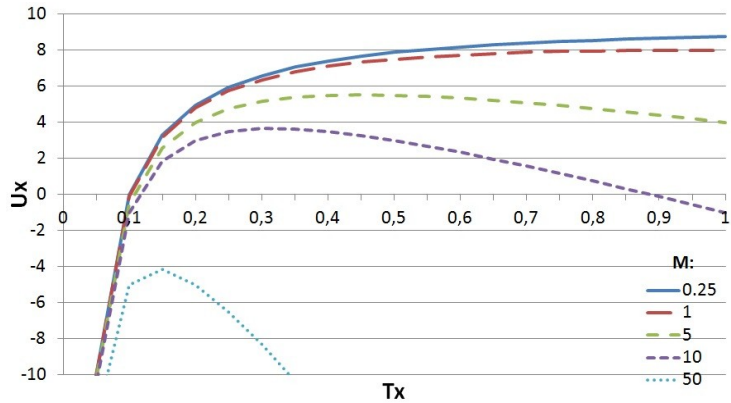

**Fig. 2.** Node *X*'s utility vs. $T_X$; several $U_X$ plots are shown for various *M* and *G* = 10.

### 5.4 Tit-for-tat and binary reputation metric

This subsection analyzes in detail equations (5) and (6) along with binary reputation metric described in sec. 4.3. Node *X*'s strategy implements a threshold $t_S$ on $T_X$ which marks a border between $R_X = 0$ and $R_X = 1$. Below this threshold no amount of requested service $S_{XN}$ is granted by the network, thus node *X*'s utility is infinite negative the node has no perception of being connected. On the other hand, nodes with $T_X \geq t_S$ get their $S_{XN}$ granted in full. Node *X*'s optimal strategy in this game variant is $T_X = t_S$, because it still can refuse to service some part of $S_{NX}$ and at the same time enjoy a full service of its requests and a full perception of being connected.

From the network *N*'s perspective, in the error-free environment ($e = 0$), the optimal value of $t_S$ would be 1, as it would force *X* to be fully cooperative and there would be no discrepancy between the proportion of service provided by *N* and *X*. However, in reality the wireless network environment is not perfect, i.e. $e > 0$, and proceeding this way would exclude some cooperative nodes from the network. The specific number of excluded nodes depends on the actual value of *e*. Given its stochastic nature, optimal selection of $t_S$ is hard.

### 5.5 Tit-for-tat and fine-grained strategy with a threshold

The last strategy examined in this paper is a modification of the tit-for-tat strategy analyzed in Section 5.3, with a threshold $t_S$ on $R_X$ above which the network *N* starts to send its service requests to node $X$. The optimal $T_X$ in this case should be just below $t_S$ in a general case. However, if $M \leq 0.5$ then the policy $T_X$ in $[t_S, 1]$ turns out to give a slightly better performance due to the lower cost of servicing $S_{XN}$.

This strategy poses the same problem of accurate definition of the threshold value as in the binary metric case. However, the discrepancy between the amount of service received and provided by the network $N$ is eliminated, as is the problem of partially cooperative nodes. The most serious issue with this strategy is the optimal $T_X$ that makes node $X$ useless to the network, and creates a strong incentive for all nodes to follow it to minimize their costs if the amount of service requested from them is significant.

## 6  Summary

We have introduced a framework for analysis of cost incentives driving nodal behavior under a reputation system, enabling comparison of different solutions and evaluation of new concepts. We have sketched some of the most common strategies of reputation systems for wireless ad hoc networks and enabled an analytical confirmation of some heuristic findings, such as the necessity of incentive-driven reputation systems in ad hoc networks, the need of indirect reciprocity towards rational nodes and the insufficiency of merely signaling uncooperative behaviors. We have showed the possibility of creating a reputation system able to shape nodes' utility functions in a way that will enforce a non-trivial service provision, by making it a strategy of highest utility for a rational node. Another contribution is made by pointing out the risk of excessive exploitation of high-reputation nodes in a way that lowers the amount of service provided to the rest of the network. Thus a need for creation of a fair load vs. reputation balancing mechanism in ad hoc networks has been shown. In the near future, the qualitative findings of this paper will be verified through extensive simulations of realistic wireless network environments and a more detailed analysis.

## 7  Acknowledgment

Preliminary ideas of the paper were developed during the Future Internet Engineering project supported by the European Regional Development Fund under Grant POIG.01.01.02-00-045/90-00.